\documentclass[aps,prb,longbibliography,twocolumn,superscriptaddress,showpacs,floatfix]{revtex4-2}
\usepackage{amsmath,amssymb,amsfonts,float,graphics,epsfig,epstopdf,color,verbatim,tabularx,bm,multirow,appendix}
\usepackage[utf8]{inputenc}
\usepackage[T1]{fontenc}
\usepackage{xcolor}
\usepackage{dsfont}
\usepackage{textcomp}
\usepackage{yfonts}
\usepackage{footnote}
\usepackage{bm}
\usepackage{subfigure}
\usepackage{mathrsfs}
\usepackage{graphicx}
\usepackage{verbatim}
\usepackage[colorlinks=true,citecolor=blue,linkcolor=red,urlcolor=blue]{hyperref}
\usepackage{multirow}
\usepackage{braket}
\usepackage[normalem]{ulem}
\usepackage{tikz}
\usetikzlibrary{calc}
\usetikzlibrary{shapes.multipart}

\DeclareMathOperator{\Tr}{Tr}

\newcommand{\comments}[1]{}

\newcommand{\Eq}[1]{Eq.~\eqref{#1}}

\newcommand{\Fig}[1]{Fig.~\ref{#1}}

\newcommand{\stkout}[1]{\ifmmode\text{\sout{\ensuremath{#1}}}\else\sout{#1}\fi}

\newcommand\startsupplement{%
       \newpage\clearpage
       \setcounter{secnumdepth}{2}
       \setcounter{table}{0}
       \renewcommand{\thetable}{S\arabic{table}}
       \setcounter{figure}{0}
       \renewcommand{\thefigure}{S\arabic{figure}}
       \setcounter{equation}{0}
       \renewcommand{\theequation}{S\arabic{equation}}
       \setcounter{section}{0}
       \renewcommand{\thesection}{Section \Roman{section}}
       \renewcommand{\thesubsection}{\Roman{section}. \Alph{subsection}}
    }

\makeatletter
\def\l@subsubsection#1#2{}
\makeatother

\graphicspath{{./Figs/}}

\begin{document}

\title{Probing Non-Fermi-Liquid Behaviour of Composite Fermi Liquid via Efficient Thermal Simulations}

\author{Bin-Bin Chen}
\affiliation{Peng Huanwu Collaborative Center for Research and Education, Beihang University, Beijing 100191, China}

\author{Hongyu Lu}
\affiliation{Department of Physics and HK Institute of Quantum Science \& Technology, The University of Hong Kong, Pokfulam Road, Hong Kong SAR, China}
\affiliation{State Key Laboratory of Optical Quantum Materials,
The University of Hong Kong, Pokfulam Road, Hong Kong SAR, China}

\author{Zi Yang Meng}
\email{zymeng@hku.hk}
\affiliation{Department of Physics and HK Institute of Quantum Science \& Technology, The University of Hong Kong, Pokfulam Road, Hong Kong SAR, China}
\affiliation{State Key Laboratory of Optical Quantum Materials,
The University of Hong Kong, Pokfulam Road, Hong Kong SAR, China}

\begin{abstract}
The physics of two-dimensional electron gas in a perpendicular magnetic field, i.e., the quantum Hall system, is remarkably rich. At half filling of the lowest Landau level, it has been predicted that ``composite fermions''---emergent quasiparticles consisting of an electron attached to two magnetic flux quanta---experience zero net magnetic field and form a Fermi sea, dubbed composite Fermi liquid (CFL). However, despite its seemingly simple appearance, CFL is a strongly correlated quantum many-body state in disguise, and solving it in a controlled manner is extremely difficult, to the extent that the thermodynamic properties of CFL remain largely unknown. In this work, we perform state-of-the-art thermal tensor network simulations of the $\nu=1/2$ Landau level system and observe low-temperature power-law behaviour of the specific heat, signaling the gapless nature of CFL. More importantly, the power is extracted to be close to $2/3$, clearly deviating from the ordinary linear-$T$ behaviour of Fermi liquid, suggesting coupling between the CFs and the dynamical emergent gauge field and thereby revealing the quantum many-body nature of the CFL state.
\end{abstract}

\date{\today}
\maketitle

\noindent{\textcolor{blue}{\it Introduction.}---}
The two-dimensional electron gas subjected to a perpendicular magnetic field—known as the quantum Hall (QH) system—exhibits 
remarkable physical richness. In the integer quantum Hall regime, the Hall conductance $\sigma_{xy}$ is quantized precisely 
in integer multiples of $e^2/h$~\cite{KlitzingPRL1980}. In the fractional quantum Hall regime, $\sigma_{xy}$ is quantized at 
fractional values of $e^2/h$, accompanied by the emergence of fractionally charged quasiparticles obeying fractional 
statistics~\cite{TsuiPRL1982,HalperinPRL1984,ArovasPRL1984}. Notably, at filling $\nu = 1/2$, the system exhibits 
metallic behaviour, characterized by a finite longitudinal resistivity {$\rho_{xx}$} 
and the absence of a plateau in $\rho_{xy}$~\cite{JiangPRB1989}.

The composite fermion (CF) theory postulates that electrons attach an even number of magnetic flux quanta, forming emergent quasiparticles that experience a reduced effective magnetic field~\cite{JainPRL1989,Jain2007composite}. Based on this flux-attachment concept, mean-field theories~\cite{ZhangPRL1989,ReadPRL1989,LopezPRB1991,HLRPRB1993} have been successful: near $\nu = 1/2$, ballistic transport experiments are consistent with composite fermions moving in a weak residual magnetic field~\cite{KangPRL1993,GoldmanPRL1994,SmetPRL1996}; at half filling, the external magnetic field is entirely canceled by the attached flux, resulting in a Fermi sea of composite fermions—a picture corroborated by surface acoustic wave measurements~\cite{WillettPRL1990}.
More recently, this conventional CF picture has been revisited in light of its connections to 3D topological insulators (TIs), reformulating composite fermions as massless Dirac fermions reminiscent of TI surface states~\cite{SonPRX2015,WangPRX2015,MrossPRX2015,MetlitskiPRB2016}.

In recent years, the fractional Chern insulator (FCI)~\cite{SunPRL2011Nearly,TangPRL2011HighTemperature,ShengNC2011Fractional,
NeupertPRL2011Fractional,RegnaultPRX2011Fractional,WangPRL2011Fractional}—a lattice analogue of the fractional quantum Hall 
effect, also known as the fractional quantum anomalous Hall (FQAH) effect in the absence of an external magnetic field—has attracted considerable interest in the context of two-dimensional moir\'e materials, with extensive theoretical 
explorations~\cite{AbouelkomsanPRR2023Quantum,LedwithPRR2020Fractional,RepellinPRR2020Chern, WuPRL2019Topological,
LiPRR2021Spontaneous,CrepelPRB2023Anomalous,Morales-DuranPRR2023Pressureenhanced,WangPRL2024Fractional,ReddyPRB2023Fractional}.
FCIs were reported in early high-magnetic-field experiments~\cite{SpantonS2018Observation} at fractional fillings of Hofstadter bands~\cite{KolPRB1993Fractional}, as well as at fractional fillings of native Chern bands 
in magic-angle twisted bilayer graphene (TBG), where a finite magnetic field was used to improve the quantum geometry, i.e., to smooth the Berry-curvature distribution and reduce violations of the trace condition of the ideal Landau-level-like relation $\Tr(g(\mathbf{k}))=|\Omega(\mathbf{k})|$~\cite{XieN2021Fractional,Parker2021Fieldtuned}.
More recently, zero-field FCIs, i.e., FQAH states, have been observed in twisted bilayer MoTe$_2$ (tMoTe$_2$) and rhombohedral graphene/hBN superlattices~\cite{CaiN2023Signatures,park_observation_2023, ZengN2023Thermodynamic, XuPRX2023, LuN2024Fractional}. 
Notably, the zero-field metallic state at $\nu=1/2$ in topological bands has also been explored both theoretically and experimentally~\cite{DongPRL2023,GoldmanPRL2023, park_observation_2023,LuN2024Fractional}.

Substantial numerical research has been dedicated to understanding the gapped FQH and zero-field FCI phases, including studies of spectral properties using the time-dependent variational principle~\cite{Kumar2022_neutral_excitation,Liu2024_geometric_FQH,Long2025_spectra} and the single-mode approximation~\cite{Repellin2014_SMA}, as well as studies of thermodynamic properties using thermal tensor network simulations~\cite{Lu2024_thermo_neutral}. In contrast, the numerical characterization of the half-filled CFL remains challenging. From this perspective, representative works include density matrix renormalization group (DMRG) investigations of the role of particle-hole symmetry~\cite{GeraedtsScience2016}, variational Monte Carlo studies of the scaling of entanglement entropy~\cite{ShaoPRL2015,MishmashPRB2016,VoineaPRB2025}, studies of the effects of anisotropy~\cite{IppolitiPRB2017Numerical}, DMRG studies of the relationship between the Fermi sea of bare electrons at zero field and that of composite fermions at high field~\cite{IppolitiPRB2017Connection}, as well as the influence of discrete rotational symmetry $C_N$~\cite{IppolitiPRB2017Composite}, and exact diagonalization calculations of the thermoelectric response~\cite{ShengPRB2020Thermoelectric}. In all these directions, new computations and discussions of often contradictory results are actively ongoing.

In this work, we investigate the thermodynamic properties of the composite Fermi liquid at half filling of the lowest Landau level using large-scale thermal tensor network simulations, specifically the tangent-space tensor renormalization group (tanTRG) method~\cite{LiPRL2023}. We observe a low-temperature specific heat that scales as $T^\alpha$, with $\alpha \approx 2/3$—indicative of gapless excitations and markedly inconsistent with the linear-$T$ dependence expected for a conventional Fermi liquid. This power-law behaviour provides direct evidence for the strongly correlated nature of the CFL state, in that the coupling between composite fermions and a dynamical emergent gauge field gives rise to non-Fermi-liquid properties. In a broader context, our results demonstrate that thermal tensor network methods provide a controlled numerical framework for exploring non-Fermi-liquid thermodynamics in strongly correlated quantum Hall systems.

\begin{figure}[t!]
	\includegraphics[width=\columnwidth]{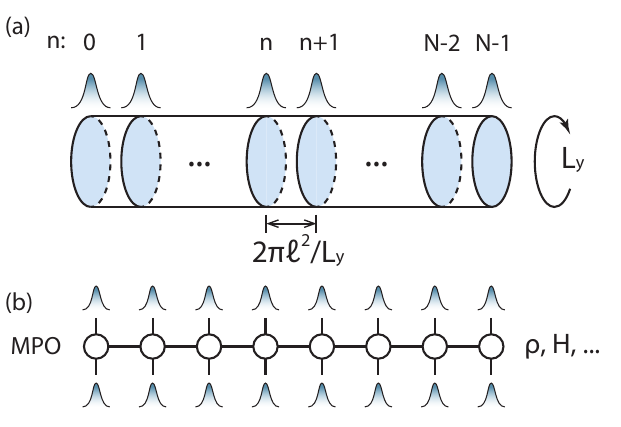}
	\caption{\textbf{The matrix product operator (MPO) setup in the lowest Landau level (LLL) basis for thermal density operator $\rho$.} 
		(a) In the Landau gauge with periodic boundary condition along the $y$-direction and open boundary condition along the $x$-direction. The LLL wavefunctions (labelled by $n \in[0,N-1]$) are extended along $y$- and localized along $x$-direction. 
		(b) The matrix product operator of density operator $\rho$ (or other operators, e.g., Hamiltonian $H$) is defined in the LLL basis, where the ``local'' degree of freedom of the $i$th tensor lives in the 2-dimensional physical space of either filling the $i$th LLL  or leaving it empty.
	}
	\label{fig:fig1}
\end{figure}

\noindent{\textcolor{blue}{\it Model definition.}---}
The model Hamiltonian we consider is defined as 
\begin{equation}
\mathcal{H} \equiv \int \mathrm{d}\mathbf{r}_1\mathrm{d}\mathbf{r}_2~ V(\mathbf{r}_1,\mathbf{r}_2)[n(\mathbf{r}_1)-\tfrac{1}{2}][n(\mathbf{r}_2)-\tfrac{1}{2}],
\end{equation}
where  $V(\mathbf{r}_1,\mathbf{r}_2)\equiv \frac{e^{-|\mathbf{r}_1-\mathbf{r}_2|/\lambda}}{ |\mathbf{r}_1 - \mathbf{r}_2|}$ is the Coulomb-Yukawa interaction between electrons, 
$n(\mathbf{r}) \equiv c^\dag(\mathbf{r})c(\mathbf{r})$ is the particle number operator, and the factor $\tfrac{1}{2}$ is introduced to maintain the half-filling condition in the 
grand-canonical ensemble. We then project the Hamiltonian onto the lowest Landau level (LLL) via $c(\mathbf{r}) = \sum_n \phi_n(\mathbf{r}) c_n$, where 
$\phi_n(\mathbf{r})\propto e^{-\frac{1}{2}(\frac{x-x_n}{\ell})^2}e^{\mathrm{i}k_n y}$ is the LLL wavefunction in cylindrical geometry (i.e., in the Landau gauge), with 
$k_n = 2\pi n/L_y$, $x_n = 2\pi n \ell^2/L_y$, $n\in[0,N-1]$, and $N$ denoting the number of LLL orbitals [c.f. \Fig{fig:fig1}(a)]. 
We set the magnetic length $\ell = \sqrt{\hbar c/(eB)}=1$ and $k_\mathrm{B}=1$ as units throughout. This yields the LLL-projected Hamiltonian, 
\begin{equation}\label{Eq:Model}
\mathcal{H}^\mathrm{LLL} = \sum_{m,n,k,l} \mathcal{A}_{m,n,k,l} c^\dag_m c^\dag_n c^{\,}_k c^{\,}_l,
\end{equation}
with the form factor
$\mathcal{A}_{m,n,k,l} =  \int \mathrm{d}\mathbf{r}_1 \mathrm{d}\mathbf{r}_2 V(\mathbf{r}_1-\mathbf{r}_2) \phi^\ast_{m}(\mathbf{r}_1)\phi_{l}(\mathbf{r}_1) \phi^\ast_{n}(\mathbf{r}_2) \phi_{k}(\mathbf{r}_2)$ evaluated numerically.

\noindent{\textcolor{blue}{\it Tangent-space tensor renormalization group (tanTRG) in the LLL basis.}---}
Here, we briefly recapitulate the idea of the tangent-space tensor renormalization group (tanTRG), 
which has proven successful in accessing the challenging low-temperature properties of quantum lattice models, including the 2D Hubbard and $t$-$J$ models for cuprates~\mbox{\cite{LiPRL2023,QuPRL2024Phase}}, the bilayer $t$-$J$-$J_\perp$ model for nickelate superconductors~\cite{QuPRL2024Bilayer}, and a real-space effective model of MoTe$_2$~\cite{chen2025fractionalcherninsulatorquantum}. 
We then show how tanTRG can be adapted to the LLL orbital basis.

To begin with, a generic Hamiltonian can be efficiently decomposed into a matrix product operator (MPO) representation~\cite{frowis_tensor_2010,hubig_generic_2017,crosswhite_applying_2008,motruk_density_2016,schollwock_density-matrix_2011}. 
At high temperature, the thermal density operator can then be constructed with high precision via a series expansion~\cite{chen_series-expansion_2017}, which effectively reduces to $\rho(\tau) \equiv e^{-\tau \mathcal{H}} = I - \tau \mathcal{H}$ for a sufficiently small inverse temperature $\tau\equiv1/T$. Here, $\tau=10^{-6}$ is taken, yielding a small expansion error of $\mathcal{O}(10^{-12})$.

We then cool down the system using a mixed temperature grid: first, an exponential grid $\beta_i=\tau\times2^{i/z}$ is used to quickly approach the low-$T$ regime, with $z=4$ chosen to ensure dense temperature points for the numerical derivative; this is followed by a linear grid in inverse temperature once the inverse-temperature increment $\Delta\beta$ exceeds a given threshold. Here, $\Delta\beta=5$ is used to ensure small projection errors in the low-$T$ regime.

The cooling step follows the flow equation 
$\partial \rho/\partial \beta = \mathcal{P}(-\mathcal{H}\rho) \mathcal{P}$, constrained to the fixed-bond-dimension matrix product operator (MPO) manifold $\mathcal{M}_\mathrm{MPO}$, 
where $-\mathcal{H}\rho$ is the tangent vector in the full Hilbert space, which generally takes the density matrix $\rho$ out of $\mathcal{M}_\mathrm{MPO}$, 
and the projector $\mathcal{P}$ serves to project the tangent vector back to the tangent space of the MPO manifold $\mathcal{T}_\rho\mathcal{M}_\mathrm{MPO}$. 
For more details on the MPO parameterization of the above constrained flow equation, we refer interested readers to Ref.~\cite{LiPRL2023}.

Specifically, in this work, we deal with the many-body basis consisting of LLL wavefunctions, 
$|q_0,q_1,\cdots,q_{N-1}\rangle = (c^\dag_0)^{q_0}(c^\dag_1)^{q_1}\cdots(c^\dag_{N-1})^{q_{N-1}}|\Omega\rangle$, 
where the quantum number $q_i$ denotes the number of particles occupying the $i$th LLL orbital, 
and $|\Omega\rangle$ is the vacuum state.
A generic operator $O$ in this many-body LLL basis can then be expressed as $O = \sum_{\{q_i, q_i'\}} O_{\{q_i\}, \{q_i'\}} |\{q_i\}\rangle \langle \{q_i'\} |$, 
where $O_{\{q_i\},\{q_i'\}} \equiv \langle \{q_i\}|O|\{q_i'\}\rangle$ is a $2^N\times2^N$-dimensional tensor and can be parameterized in MPO form as
$O_{\{q_i\},\{q_i'\}} = \sum_{b_0, b_1, \cdots, b_N} \prod_{i=0}^{N-1} A_{b_i, b_{i+1}, q_i, q_i'}$, with 
the virtual indices $b_i$ ranging from $1$ to $D$ (the bond dimension), as shown in Fig.~\ref{fig:fig1}(b). 
For the projected Hamiltonian $\mathcal{H}^\mathrm{LLL}$ and the initial high-temperature density operator $\rho(\tau)$, 
we can find their exact MPO forms, and the lower-temperature density operator $\rho(\beta)$ can then be tracked using the tanTRG protocol described above. We note that our method can be straightforwardly generalized to FQAH and quantum moir\'e systems~\cite{zhangMomentum2021,AbouelkomsanPRR2023Quantum,LedwithPRR2020Fractional,RepellinPRR2020Chern, WuPRL2019Topological,
LiPRR2021Spontaneous,CrepelPRB2023Anomalous,Morales-DuranPRR2023Pressureenhanced,WangPRL2024Fractional,ReddyPRB2023Fractional,huangAngle2025} and other projected-Hamiltonian systems~\cite{wangPhases2021,chenPhases2024}, where one only needs to change the basis functions onto which the interaction is projected.

\begin{figure}[t!]
	\includegraphics[width=\columnwidth]{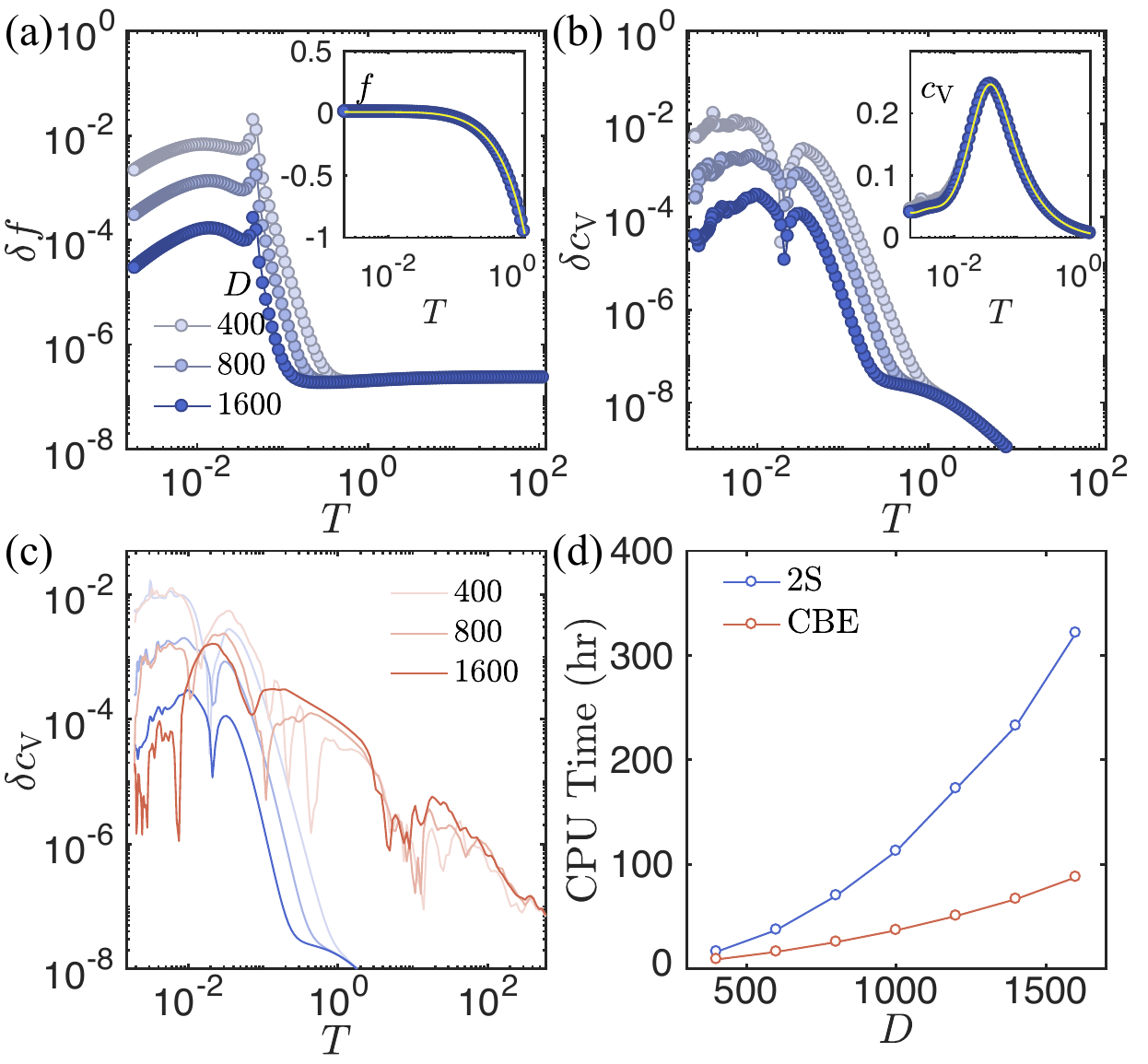}
	\caption{\textbf{Benchmark and CBE optimization of tanTRG for the projected Coulomb-Yukawa interaction.} We choose $N=16$ and compare to the exact diagonalization results. 
	(a) Relative errors of free energy $\delta f \equiv |f_\mathrm{tanTRG} - f_\mathrm{ED}|/|f_\mathrm{ED}|$ calculated with different bond dimensions $D=400, 800, 1200$ are shown versus temperature $T$. (b) Relative errors of specific heat $\delta c_\mathrm{V}$ are shown versus temperature $T$. 
	In the inset of both panels, the raw tanTRG data of the free energy $f$ and specific heat $c_\mathrm{V}$ is shown along with the ED data (the yellow curves). 
(c) Relative errors of specific heat $\delta c_\mathrm{V}$ for both 2-site {(blue, the same as in (a)) and CBE (red)}  update scheme with various bond dimensions $D=400, 800, 1600$. (d) The CPU time (in the unit of hours) for both 2-site and CBE update scheme, shown as functions of bond dimension $D$.}
	\label{fig:fig2}
\end{figure}

\noindent{\textcolor{blue}{\it Benchmark results.}---}
In \Fig{fig:fig2}, we benchmark tanTRG for a small system size, $N=16$. Here, we take the circumference of the cylinder to be $L_y=12$ and the effective 
screening length to be $\lambda=10\tfrac{2\pi\ell^2}{L_y}$. We first calculate the free energy per orbital, $f(T) \equiv -\frac{1}{N}T\ln Z$, with $Z = \Tr\rho$ being the partition function. 

In \Fig{fig:fig2}~(a), we show three curves of the relative errors in the free energy, $\delta f\equiv |f_\mathrm{tanTRG}-f_\mathrm{ED}|/|f_\mathrm{ED}|$, for increasing bond dimensions $D=400, 800, 1600$. For all three curves, in the high-temperature regime, the relative errors remain within the order of $10^{-6}$, as the discarded weights are negligible and the final error comes from the Trotter error and the projection error, the latter resulting from projecting the full tangent vector into $T_\rho\mathcal{M}_\mathrm{MPO}$. 
At intermediate temperatures, $T\sim10^{-1}$, the thermal state changes significantly, as indicated by the round peak in the specific heat [c.f. the inset of \Fig{fig:fig2}~(b)], 
and finite discarded weights set in, resulting in relatively larger errors in the free energies.
In the lower-temperature regime, the finite-size gap sets in, and the thermal density operator $\rho$ converges to $|\psi_0\rangle\langle\psi_0|$ (the ground state), which in this example possesses 
lower (purified) entanglement entropy and thus lower $\delta f$ than in the intermediate-$T$ cases. 
Such lower-$T$ behaviour can also be understood from the MPO-manifold perspective: the manifold becomes flattened when the finite-size gap sets in, yielding small projection errors.
Furthermore, by increasing $D$, the precision improves from $10^{-3}$ to $10^{-5}$ in the low-$T$ regime.

In \Fig{fig:fig2}~(b), we present the relative error of the specific heat, $c_V \equiv \partial e/\partial T$, with $e\equiv \frac{1}{N} \Tr(\mathcal{H}\rho)/Z$ being the internal energy per orbital. 
Similar to the free energy, the specific heat $c_V$ is obtained with high precision, with relative errors within $10^{-8}$ at high temperatures $T\gtrsim1$. These errors increase quickly at $T\sim10^{-1}$, 
while at lower temperatures $\delta c_V$ even decreases slightly due to the presence of the finite-size gap. Furthermore, by increasing the bond dimension from $D=400$ to $D=1600$, the 
precision again improves from $10^{-2}$ to $10^{-4}$.

The computational complexity can be further reduced by replacing the 2-site (2S) update scheme with the recently proposed controlled bond expansion (CBE)~\cite{GleisPRL2023,LiPRL2024},
which has been shown to improve computational efficiency while maintaining precision similar to that of the 2S scheme in DMRG and TDVP simulations. 
Theoretically, this results in a speedup by a factor of $d$ (here $d=2$) in MPS-based algorithms and by a factor of $d^2$ in MPO-based algorithms, which is very promising for our 
finite-$T$ simulations.

In \Fig{fig:fig2}~(c), we show the relative errors of the specific heat with respect to the ED results. It can be seen that, in the low-temperature regime $T\lesssim10^{-1}$, 
the 2S and CBE results exhibit similar accuracies for various bond dimensions, $D=400, 800$, and $1600$, which demonstrates the validity of CBE for the projected Hamiltonian considered here.
In \Fig{fig:fig2}~(d), we compare the computational costs of these two types of simulations, namely 2S and CBE. Overall, the CBE simulations consume much less CPU time than 
the 2S simulations. At small bond dimensions, due to numerical overhead, the acceleration of the CBE scheme is below the ideal factor of $d^2=4$. However, as the bond dimension increases, 
the speedup achieved by CBE becomes increasingly pronounced; for example, at the largest bond dimension considered, $D=1600$, the acceleration exceeds a factor of $3$. Overall, we find that tanTRG combined with the CBE update provides an efficient and high-precision approach to studying the finite-temperature properties of LLL-projected Hamiltonians.

\begin{figure}[t!]
	\includegraphics[width=\columnwidth]{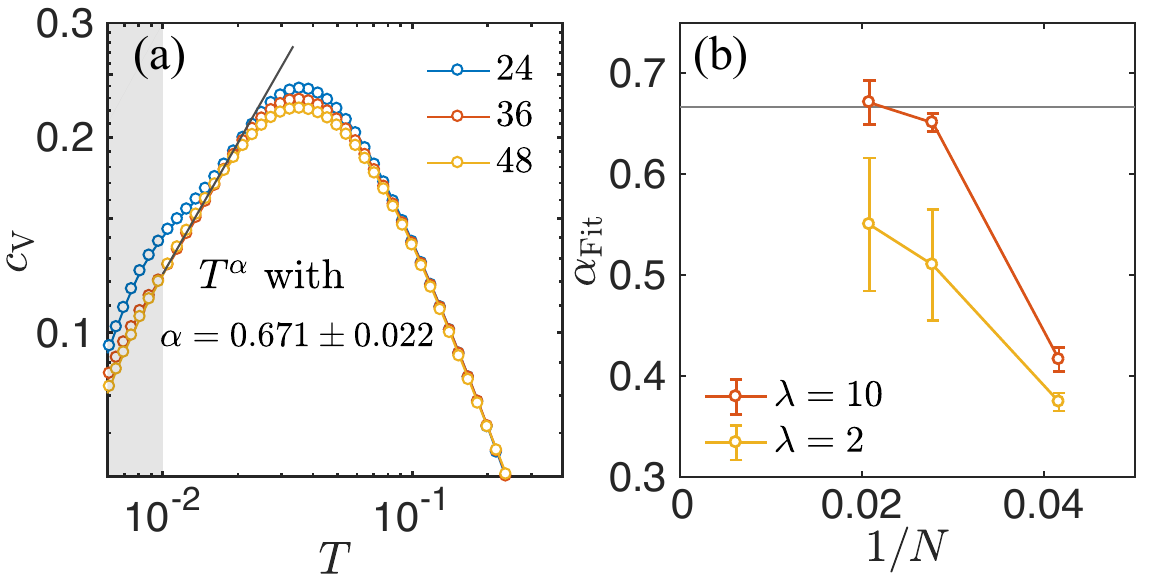}
	\caption{\textbf{Thermodynamic properties of composite Fermi liquid.} 
        (a) Lower-temperature specific heat $c_\mathrm{V}$ behaviour for various number of orbitals $N=24, 36, 48$, 
        exhibiting converged power-law scaling $T^{2/3}$ for $N=36$ and $48$. 
        (b) The low-$T$ power $\alpha_\mathrm{Fit}$ from a linear fitting is shown versus system sizes $1/N$ for $\lambda=10$ and $2$.
	}
	\label{fig:fig3}
\end{figure}

\begin{figure}[t!]
	\includegraphics[width=\columnwidth]{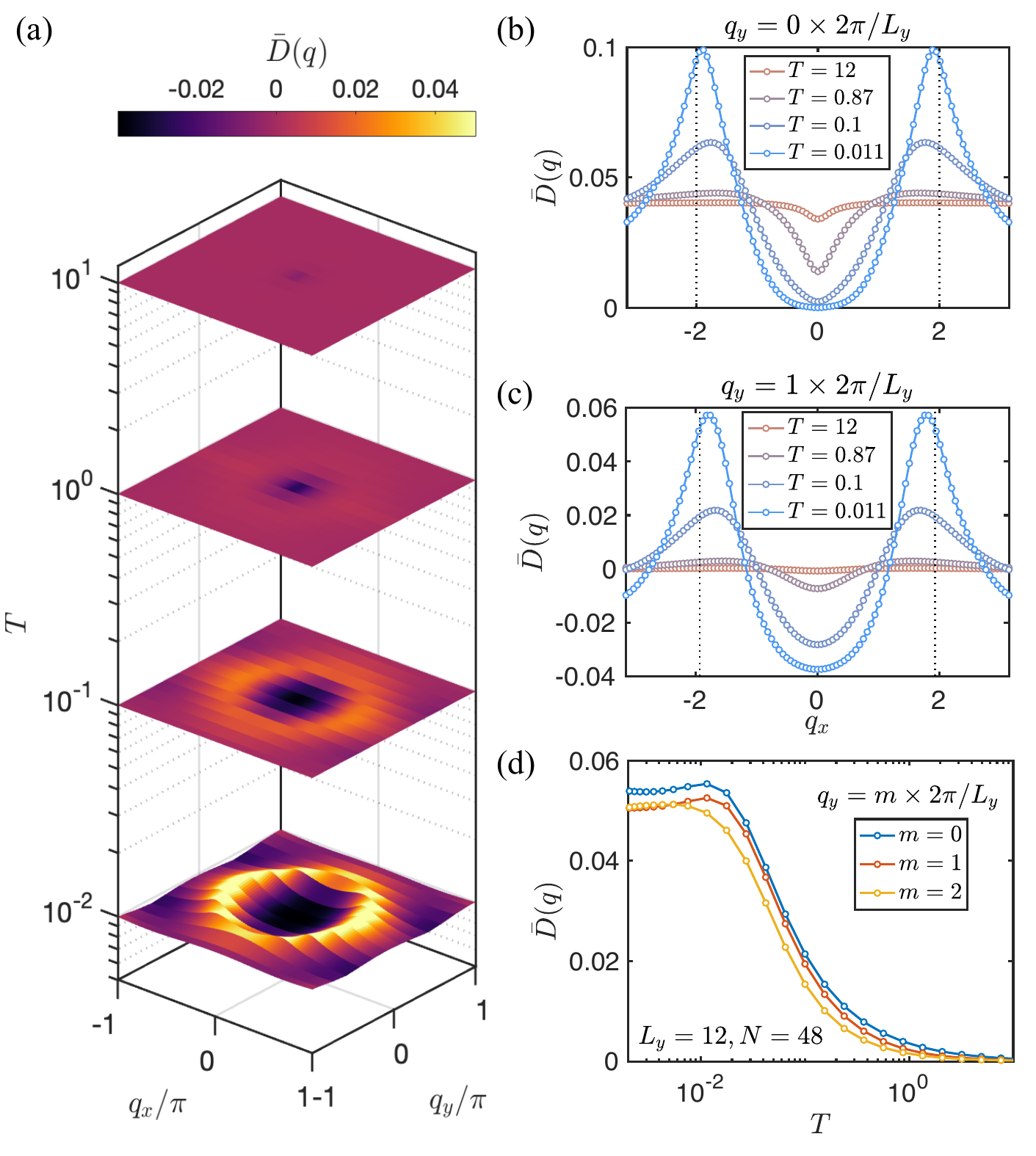}
	\caption{\textbf{Formation of the composite Fermi surface in CFL.} 
	For the system with $L_y=12, N=48,\lambda=10$,
        (a) Two-dimensional momentum space view of the ``guiding-center'' density-density correlations $\bar D(q) =\exp(q^2/2) \langle n_q n_{-q}\rangle_T$ is 
        plotted as a function of 2D momentum $q=(q_x,q_y)$ at four different temperatures $T=10^1, 10^0, 10^{-1}, 10^{-2}$. 
        It starts to exhibit a $2k_F$ circle for $T\lesssim10^{-1}$ suggesting the establishment of the composite Fermi surface. 
        (b,c) $\bar D(q)$ is shown as functions of $q_x$ 
        when $q_y=0, \frac{2\pi}{L_y}$ are fixed accordingly. The dashed grey line indicated the $2k_F$ points for the corresponding $q_y$. 
        (d) $\bar D(q)$ at the three different $2k_F$ points of $q_y=m\frac{2\pi}{L_y}$ with $m=0,1,2$, which increase most quickly at $\sim10^{-1}$, 
        suggesting the formation of the composite Fermi surface.
	}
	\label{fig:fig4}
\end{figure}

\noindent{\textcolor{blue}{\it Finite-temperature properties of Composite Fermi Liquid.}---}
We then move on to larger system sizes with various numbers of orbitals, $N=24, 36, 48$, at a fixed circumference $L_y=12$ and a relatively long screening length $\lambda=10$. 
As shown in \Fig{fig:fig3}~(a), we calculate the specific heat $c_{V}$ down to low temperatures $T\lesssim10^{-2}$. 
For the following analysis, we exclude the lower-$T$ data, as indicated by the grey area, to minimize finite-size effects. This is suggested by the compressibility 
$\partial \langle n\rangle /\partial \mu$ data [c.f. SM~\cite{suppl}],
 which quickly vanishes at $T\sim10^{-2}$, indicating a finite-circumference gap.
Above the grey regime, the three specific heat curves exhibit power-law behaviour, $T^\alpha$, and all of them clearly deviate from the linear-$T$ Fermi-liquid behaviour. As shown in \Fig{fig:fig3}~(b), notably, for the two larger sizes, $N=36$ and $48$, the power $\alpha$ converges to the value $2/3$, 
consistent with the value predicted by Halperin, Lee, and Read (HLR)~\cite{HLRPRB1993}. In the plot, we also show the data for the smaller screening length $\lambda=2$; 
these data seem to suffer from stronger finite-size effects, while the power still tends to approach $2/3$. Such a non-Fermi-liquid thermodynamic response, distinct from the $C\sim T$ behavior of a normal metal, provides decisive evidence for strong interactions among the CFs. These interactions are mediated by the emergent gauge field in the QED$_3$ description~\cite{HLRPRB1993,SonPRX2015,WangPRX2015,MrossPRX2015,MetlitskiPRB2016,leeLowenergy2009, XYXuU12018}, or by ferromagnetic and nematic critical bosonic fluctuations~\cite{metlitskiQuantum2010,xuNonFermi2017,xuIdentification2020}.

We can also demonstrate the formation of the CFL state during the cooling process. 
In the 2D limit, the composite fermions in CFL are expected to form a circular Fermi surface with Fermi momentum $k_F = 1$. 
However, this is not manifested in the spectral function of the bare electrons. Rather, due to the non-Fermi-liquid nature of the state, it can be verified from the ``guiding-center'' density-density correlation 
$\bar D(q) = \exp(q^2/2) \langle n_q n_{-q}\rangle_T = \exp(q^2/2)  \Tr[n_q n_{-q} \exp(-\mathcal{H}/T)]/Z$, which 
reflects scattering events of CFs near the composite Fermi surface through a $2k_F$ singularity circle in momentum space. 
In \Fig{fig:fig4}~(a), we plot $\bar D(q)$ as a function of the 2D momentum 
$q=(q_x,q_y)$ at four different temperatures, $T=10^1, 10^0, 10^{-1}, 10^{-2}$. It starts to exhibit a $2k_F$ circle for $T\lesssim10^{-1}$, 
providing evidence for the formation of the composite Fermi surface. 
As shown in panels (b) and (c), we choose two fixed-$q_y$ cuts, specifically $q_y = 0$ and $2\pi/L_y$, in the 2D momentum contour. 
The dashed grey lines indicate the $2k_F$ points for the corresponding $q_y$ values. One can then see a clear broad peak around the $2k_F$ points, 
which become true singularity points in the 2D limit; these broad peaks build up significantly around $T\sim 10^{-1}$, consistent with our finding above. 
This can also be seen in panel (d), where $\bar D(q)$ at the three different $2k_F$ points (with $q_y=m\frac{2\pi}{L_y}$ and $m=0,1,2$) 
increases most rapidly at $T\sim10^{-1}$. Such a temperature scale matches that of the non-Fermi-liquid specific heat observed in Fig.~\ref{fig:fig3}.
We note that, in the low-temperature regime, the small-$q$ density correlation $D(q)$ shows Fermi-liquid-like $q^3$ behavior 
[c.f. SM~\cite{suppl}]. This implies that the CFs themselves, to some extent, behave like electrons in a conventional Fermi liquid, and that it is their strong coupling to the emergent gauge field 
that contributes to the non-Fermi-liquid behavior of the specific heat.
 
\noindent{\textcolor{blue}{\it Conclusion.}---}
The results in Figs.~\ref{fig:fig3} and \ref{fig:fig4} clearly show that the CFL is a metallic state but not a Fermi liquid. Specifically, the temperature dependence of the specific heat $(\sim T^{2/3})$ is strongly renormalized by interactions among the CFs, mediated by the emergent gauge field. Moreover, the Fermi surface of the electrons is absent; instead, the CFs are organized only into particle-hole bound states with a gapless response at $2k_F$~\cite{HLRPRB1993}, which can be detected by the density-density correlation reflecting scattering events of CFs.
Such a non-Fermi-liquid state closely resonates with those emerging from Fermi surfaces coupled to a critical $U(1)$ gauge field~\cite{leeLowenergy2009,XYXuU12018}, ferromagnetic bosonic fluctuations~\cite{xuNonFermi2017,xuIdentification2020}, and nematic fluctuations~\cite{metlitskiQuantum2010}. Previous numerical results in the same setting, either at the ground state~\cite{GeraedtsScience2016} or at finite temperature but with small-size ED~\cite{ShengPRB2020Thermoelectric}, have hinted at these properties. Nevertheless, it is in the present work that an explicit demonstration of the thermodynamic formation of the CFL non-Fermi liquid has been achieved through large-scale thermal tensor network simulations~\cite{LiPRL2023}. This is similar to the Dirac spin liquid state of spinons coupled to a gauge field in condensed matter~\cite{Hermele2004,Hermele2005,Assaad2005,ranProjected2007,ranSpontaneous2009,XYXuU12018,chenEmergent2025,fengScalable2025} and the strongly coupled conformal field theory of QED$_3$ in high-energy physics~\cite{Fiebig1990,Herbut2003,Fiore2005,Armour2011,karthikNumerical2019,di2017scaling,chester2016towards,Albayrak2022}.

Furthermore, our findings demonstrate the efficacy of the thermal tensor network approach within the framework of projected Hamiltonians~\cite{wangPhases2021,chenPhases2024}, particularly in momentum space. This highlights its potential utility for continuum models in other FQH systems~\cite{wangHybrid2026} as well as FQAH and quantum moir\'e materials, where one only needs to change the basis functions, to study thermodynamics, magnetism, and topology~\cite{Dong2024_hall_crystal, Liu2024_AFM_CI, Goncalves2025_FCI_excitation}. 
For example, our methodology can be straightforwardly applied to the zero-field CFL in twisted TMD systems~\cite{DongPRL2023,GoldmanPRL2023, park_observation_2023,LuN2024Fractional} to explore whether such a sublinear-$T$ specific heat persists in the absence of Landau levels. This might offer a thermal signature that distinguishes the CFL from a Fermi liquid in quantum moir\'e experiments.

\begin{acknowledgments}
{\it{Acknowledgment.-}} We thank Dung Xuan Nguyen, Bo Yang and Duncan Haldane for discussion on the subject. We acknowledge the support from the Research Grants Council (RGC) of
Hong Kong (Project Nos. 17309822, C7037-22GF, 17302223, 17301924, 17301725), the ANR/RGC Joint Research Scheme sponsored by RGC of Hong Kong and French National Research Agency (Project No. A\_HKU703/22). We thank HPC2021 system under the Information Technology Services at the University of Hong
Kong~\cite{hpc2021}, as well as the Beijing Paratera Tech Corp., Ltd~\cite{paratera} for providing HPC resources that have contributed to the research results
reported within this paper.
\end{acknowledgments}

\bibliography{bibtex}

\startsupplement

\begin{widetext}
\begin{center}
{\bf \uppercase{Supplementary Materials for \\[0.5em]
Probing Non-Fermi-Liquid Behaviour of Composite Fermi Liquid via Efficient Thermal Simulations}}

\end{center}

\vskip3em

In Supplementary Materials \ref{sec:I}, we provide detailed derivation of a generic interaction (e.g. Coulomb-Yukawa interaction considered in the main text) projected onto the lowest Landau level on a cylinder.
In \ref{sec:II}, we provided more detailed thermodynamic data of composite Fermi Liquid at the half filling of lowest Landau level.

\section{Generic Interaction projected onto LLL basis}\label{sec:I}

We consider a generic two-body  interaction $V(r_1-r_2)$,
\begin{align}
\mathcal{H} &= \int dr_1 \int dr_2 V(r_1-r_2):\hat\rho(r_1)\hat\rho(r_2):  \\
&=\int dr_1 \int dr_2 V(r_1-r_2):c^\dag(r_1)c(r_1) c^\dag(r_2) c(r_2):
\end{align}
where $c(r)$ is a spinless fermionic annihilation operator on position $r$.
We then project it onto the lowest Landau level on a $L_x\times L_y$ cylinder 
(periodic boundary condition along $y$ axis), i.e., 
\begin{equation}
c(r) = \sum_n \phi_n(r) c_n
\end{equation}
Here, we take the Landau gauge $(A_x,A_y)=(0,Bx)$, and the wavefunction takes the form of
\begin{equation}
\phi_n(r) = \frac{1}{\pi^{1/4}\sqrt{L_y \ell}} e^{-\frac{1}{2}(\frac{x-x_n}{\ell})^2} e^{i k_n y},
\end{equation}
with $k_n=\frac{2\pi n}{L_y}$, $x_n = \frac{2\pi n}{L_y} \ell^2$, and the magnetic length $\ell=\sqrt{\hbar c/(eB)}$.
We then arrive at 
\begin{align}
\mathcal{H} &= \sum_{n_1,m_1,n_2,m_2} \int dr_1 \int dr_2 V(r_1-r_2) \phi^\ast_{n_1}(r_1)\phi_{m_1}(r_1) \phi^\ast_{n_2}(r_2) \phi_{m_2}(r_2)
:c^\dag_{n_1} c_{m_1} c^\dag_{n_2} c_{m_2}: \\
&= \sum_{n_1,m_1,n_2,m_2} \mathcal{A}_{n_1,n_2,m_2,m_1}
:c^\dag_{n_1} c_{m_1} c^\dag_{n_2} c_{m_2}: \\
\end{align}
For the form factor, we have
\begin{equation}
A_{n_1,n_2,m_2,m_1} =  \int dr_1 \int dr_2 V(r_1-r_2) \phi^\ast_{n_1}(r_1)\phi_{m_1}(r_1) \phi^\ast_{n_2}(r_2) \phi_{m_2}(r_2).
\end{equation}
For the Yukawa interaction
\begin{equation}
V(r) = \frac{e^2}{4\pi\epsilon}\frac{1}{r} \exp(-r/\lambda),
\end{equation}
with the Fourier transformation
\begin{equation}
\tilde V(Q) = \frac{e^2}{4\pi\epsilon}\frac{1}{\sqrt{Q^2+1/\lambda^2}} = \frac{e^2}{4\pi\epsilon} \frac{1}{\sqrt{q_x^2 + q_y^2 +1/\lambda^2}}.
\end{equation}
The form factor matrix element will then be
\begin{align}
\mathcal{A}_{n_1,n_2,n_3,n_4} =& 2 \frac{e^2}{4\pi\epsilon} \frac{\delta_{n_1+n_2,n_3+n_4}}{L_y} \exp\left(-\frac{1}{2}(k_1-k_4)^2\right) \int_0^\infty dq_x\frac{\cos(q_x(k_1-k_3))}{\sqrt{q_x^2+(k_1-k_4)^2+1/\lambda^2}} 
\exp\left(-\frac{1}{2}q_x^2\right).
\end{align}

\section{More data on the thermodynamics of CFL}\label{sec:II}

In this section, we first examine the convergence of the specific-heat data obtained from our tanTRG simulations. As shown in \Fig{fig:ConvCheck}(a), the specific heat $c_V$ is plotted as a function of temperature $T$ for $N=24$, $\lambda=10$, and various bond dimensions $D=800, 1200, 1600$. We note that $c_V$ is well converged above the grey area where finite-size effects set in, as discussed below. This validates the extraction of the power $\alpha$ for the low-$T$ specific-heat behavior $c_V\sim T^\alpha$. For larger system sizes, $N=36$ and $48$, we show in panels (b) and (c) that $c_V$ is also well converged in the region discussed above.

\begin{figure}[t!]
	\includegraphics[width=.95\columnwidth]{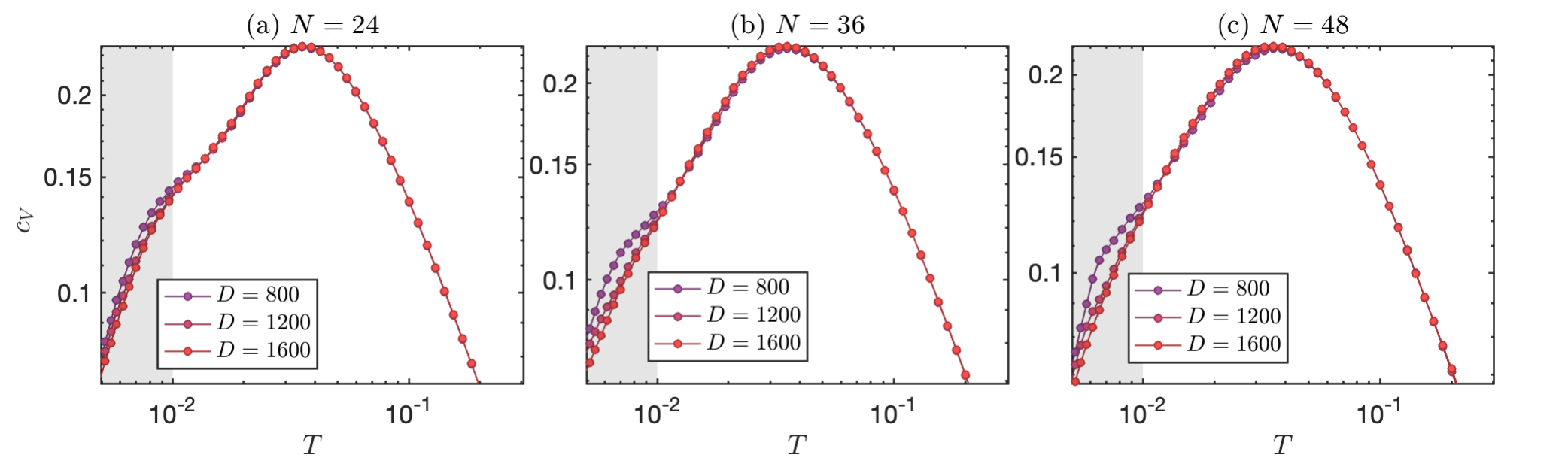}
	\caption{
        (a) Specific heat data obtained from tanTRG calculations with various bond dimensions ranging from $D=800$, $1200$ to $1600$, for half-filled Landau-level system with number of orbitals $N=24$, and screening length $\lambda=10$. (b) Similar data as panel (a) with $N=36$ otherwise. (c) Similar data as panel (a) with $N=48$ otherwise. 
	}
	\label{fig:ConvCheck}
\end{figure}

We next perform a direct comparison between the power-law and $T\log(T)$ fits on the same data set. As shown in \Fig{fig:fitting}(a), we fit the lowest five data points above the finite-size scale, indicated by the filled blue dots, using both a power law, $\sim T^\alpha$, and the $-T\log(T)$ behavior. {\bf The power-law fit yields a visibly smaller residual.} In \Fig{fig:fitting}(b), we also plot $-c_V/\log(T)$ versus $T$; a pure $T\log T$ behavior would appear as a straight line, whereas our data show a clear deviation from linearity.

\begin{figure}[t!]
	\includegraphics[width=.75\columnwidth]{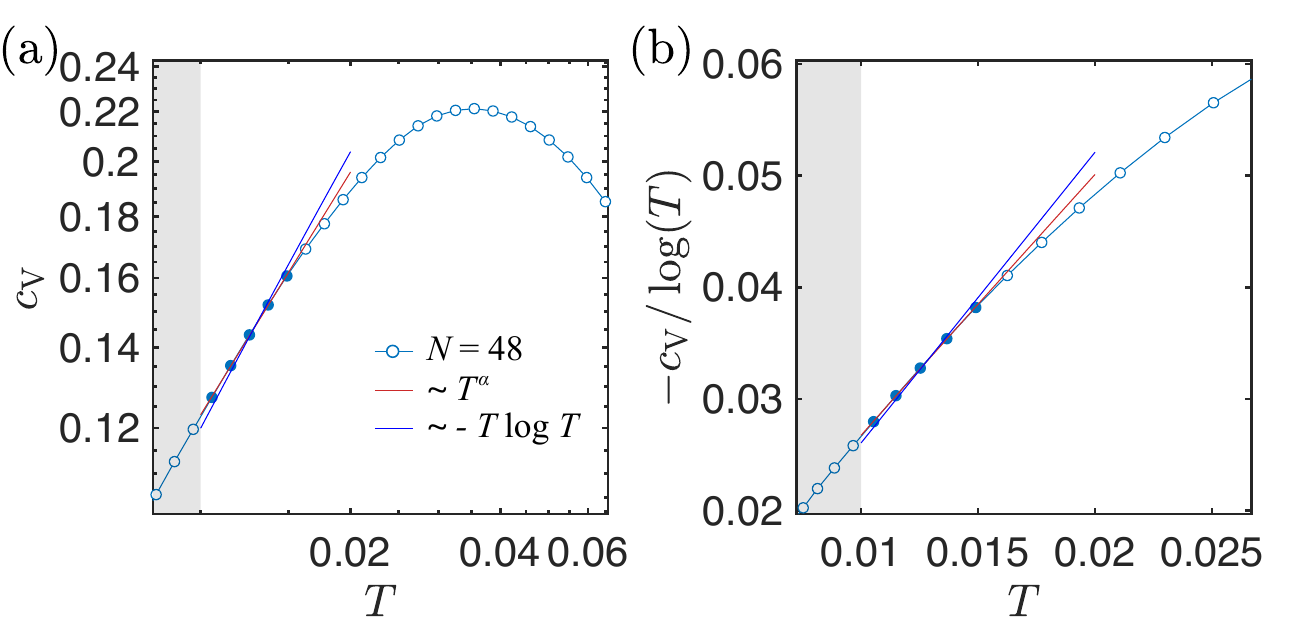}
	\caption{
        (a) Lower-temperature specific heat $c_\mathrm{V}$ for $N=48$ and $\lambda=10$ as function of temperature $T$. 
	(b) $-c_\mathrm{V}/\log(T)$ for $N=48$ as function of temperature $T$. 
	}
	\label{fig:fitting}
\end{figure}

\begin{figure}[t!]
	\includegraphics[width=0.5\columnwidth]{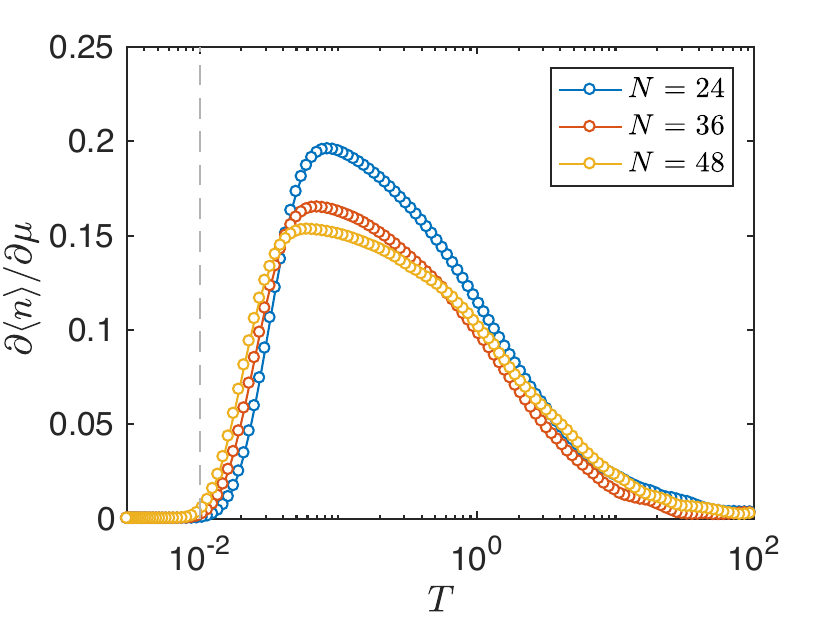}
	\caption{Charge compressibility $\partial \langle n\rangle /\partial \mu$ is shown versus $T$, which quickly drops to zero at the low-$T$ regime 
        (denoted by the grey area) suggesting the finite-size effect.
	}
	\label{fig:figsm1}
\end{figure}

\begin{figure}[t!]
	\includegraphics[width=0.8\columnwidth]{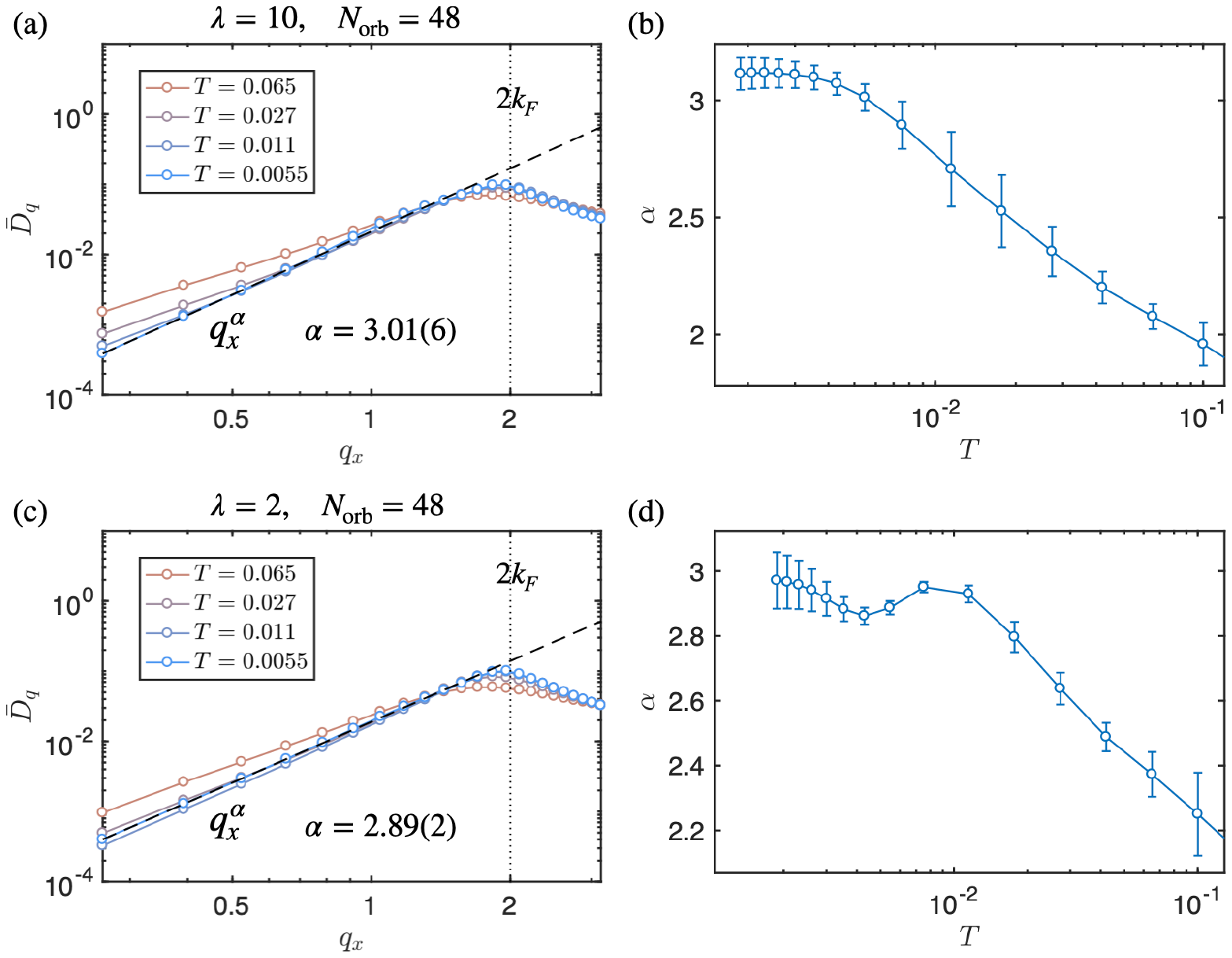}
	\caption{Guiding-center density correlation $\bar D(q)$ data at low-$T$ regime. 
	(a) For $\lambda=10$ and $N=48$, $\bar D(q)$ is shown versus $q_x$ at fixed $q_y=0$, showing power-law behaviour $q_x^\alpha$ in low-$T$ and small-$q$ regime.
	At relatively low temperature $T=0.0055$, the extracted $\alpha$ is $3.01(6)$, close to the Fermi-liquid behaviour $q^3$. 
	(b) The extracted $\alpha_\mathrm{Fit}$ is shown versus temperature $T$, which converges to $3$ at low-$T$ regime within the errorbar of the linear fitting.
	(c,d) Similar plots as (a,b) with smaller $\lambda=2$, where the low-$T$ behaviour is slightly deviated from Fermi-liquid behaviour $q^3$.}
	\label{fig:figsm2}
\end{figure}

As shown in \Fig{fig:figsm1}, for fixed circumference $L_y=12$ and $\lambda=10$, the charge compressibility $\partial \langle n\rangle/\partial \mu$ is calculated by 
including a chemical-potential term $\mu\sum_n \hat\rho_n$ in the model Hamiltonian \Eq{Eq:Model}, and thus by performing grand-canonical-ensemble simulations. 
The obtained charge compressibility is shown to vanish rapidly at $T\sim10^{-2}$, signaling that the system enters an incompressible state due to finite-size effects. 
Below $T\sim10^{-2}$, the thermodynamic data, e.g., the specific-heat data, are therefore excluded from the analysis in the main text.

As shown in \Fig{fig:figsm2}, we have calculated the guiding-center density correlation $\bar D(q)$ in the low-$T$ regime. 
In panel (a), for $\lambda=10$ and $N=48$, $\bar D(q)$ is shown as a function of $q_x$ at fixed $q_y=0$, exhibiting power-law behavior $q_x^\alpha$ in the low-$T$ and small-$q$ regime.
At the relatively low temperature $T=0.0055$, the extracted $\alpha$ is $3.01(6)$, close to the Fermi-liquid behavior $q^3$. 
In panel (b), the extracted $\alpha_\mathrm{Fit}$ is shown as a function of temperature $T$, and it converges to $3$ in the low-$T$ regime within the error bar of the linear fit.
As shown in panels (c) and (d), for a smaller $\lambda=2$, similar to the specific-heat behavior, the data seem to suffer from stronger finite-size effects, as the extracted power $\alpha$ oscillates 
around $3$ in the low-$T$ regime.

\end{widetext}
\end{document}